\documentclass[aps,prd,amssymb,showpacs,floatfix,preprint,nofootinbib,superscriptaddress]{revtex4-1}
\usepackage{upgreek,amsmath,amsfonts,graphics,epsfig,amssymb,color}

\definecolor{DarkGreen}{RGB}{0,0,127}

\newcommand{\black}{\protect\color{black}}

\begin{document}

\title{Muon content of extensive air showers: comparison of
the energy spectra obtained by the Sydney University Giant Air-shower
Recorder and by the Pierre Auger Observatory}

\author{J.\,A.\,Bellido}
\author{R.\,W.\,Clay}
\affiliation{Physics Department, The University of Adelaide, SA-5005, Australia}

\author{N.\,N.\,Kalmykov}
\affiliation{D.V.~Skobeltsyn Institute of Nuclear Physics,
M.V.~Lomonosov Moscow State University, Moscow 119991, Russia}

\author{I.\,S.\,Karpikov}
\email{karpikov@inr.ru (corresponding author)}
\author{G.\,I.\,Rubtsov}
\author{S.\,V.\,Troitsky}
\affiliation{Institute for Nuclear
Research of the Russian Academy of Sciences, 60th October Anniversary
Prospect 7a, Moscow 117312, Russia}

\author{J.\,Ulrichs}
\affiliation{School of Physics, University of Sydney, NSW-2006, Australia}
%\author{B.\,R.\,Dawson}
%\affiliation{Physics Department, The University of Adelaide, SA-5005,
%Australia}

%\pacs{95.85.Ry, 98.70.Sa, 98.70.Vc.}

%\preprint{}
%\date{}

\begin{center}
\begin{abstract}
The Sydney University Giant Air-shower Recorder (SUGAR) measured the
 energy spectrum of ultra-high-energy cosmic rays reconstructed from
muon-detector readings, while the Pierre Auger Observatory, looking at
the same Southern sky, uses the calorimetric fluorescence method for the
same purpose. Comparison of their two spectra allows us to reconstruct the
empirical dependence of the number of muons  in a vertical
shower on the primary energy for energies between $10^{17}$ and
$10^{18.5}$~eV. We compare this dependence with the predictions of
hadronic interaction models \mbox{QGSJET-II-04}, \mbox{EPOS-LHC}  and
SIBYLL-2.3c. The empirically determined number of muons with
energies above 0.75~GeV in  a vertical shower exceeds the
simulated one by the factors $\sim$1.7 and $\sim$1.3 for $10^{17}$~eV
proton and iron primaries, respectively. The muon excess grows moderately
with the primary energy, increasing by an additional factor of $\sim 1.2$
for $10^{18.5}$~eV primaries.
\end{abstract}
\end{center}
\maketitle

\section{Introduction}
\label{sec:intro}
In recent years, considerable attention has been paid to
discrepancies between theoretical models of the development of extensive
air showers (EAS), realized in Monte-Carlo simulations, and real EAS data,
see e.g.\ Ref.~\cite{Kampert_and_Unger} for a review and
Ref.~\cite{EAS-MSU_mu} for a recent summary. Understanding of
fundamental reasons for these discrepancies might shed light on the
physics of hadronic interactions at energies and in kinematical
regions hardly accessible in accelerator experiments. One of the best
studied, though not yet understood, discrepancies is the so-called muon
excess. Analyses of the muon content of air showers initiated by primary
particles with energies $\gtrsim 10^{19}$~eV by the Pierre Auger
Observatory \cite{PAOmu-old, PAOmu1, PAOmu2} and the Yakutsk
experiment~\cite{Yak-mu} revealed an overall excess (several tens per
cent, depending on the assumed primary composition and interaction model)
in the number of muons with energies $E_{\mu} \gtrsim 1$~GeV in EAS,
compared to simulations performed with available hadronic-interaction
models. This is in line with earlier results of the HiRes/MIA
experiment~\cite{HiRes-MIA} obtained for $E \gtrsim 10^{17}$~eV. While
preliminary IceTop results~\cite{IceTop} for GeV muons and
$10^{15}$~eV$\lesssim E \lesssim 10^{17}$~eV suggested that no excess is
seen, they have been superseded by newer preliminary
results~\cite{IceTop_v2} demonstrating the rise of the muon excess near
$10^{17}$ eV, consistent with the excess seen by HiRes/MIA (see also
discussion in Ref.~\cite{Watson}). The NEVOD/DECOR group has studied the
number, multiplicity and energy of bundles of  $E_{\mu } \gtrsim 2$~GeV
muons in inclined showers at $E\sim10^{17}$~eV and found the excess of the
number of bundles over the simulation \cite{NEVOD} while the energy
deposited per bundles agrees well with models \cite{NEVOD_v2}.  Analysis
of
the muon density of EAS in KASCADE-Grande~\cite{KASCADE-Grande} for  muons
with $E_{\mu }\gtrsim 230$~MeV and $10^{16}$~eV$\lesssim E \lesssim
10^{17}$~eV did not reveal the excess in the muon density though a mismatch
in the muon attenuation length was found. Also, the analysis of the muon
data of the EAS-MSU experiment \cite{EAS-MSU_mu} did not reveal any excess
of $E_{\mu } \gtrsim 10$~GeV muons at $10^{17}$~eV$\lesssim E \lesssim
10^{18}$~eV. As it is discussed in Ref.~\cite{EAS-MSU_mu}, all these data
have been obtained in different conditions, that is for different ranges
of the primary energy $E$, various muon energies $E_{\mu}$, at different
atmospheric depths, for different zenith angles and with
data obtained at different distances from the shower axis. A more
systematic study of the muon excess over a large range of
primary energies and zenith angle is needed with a single experiment.
%Unfortunately, currently operating EAS
%experiments have only a limited number of muon detectors: 6 detectors over
%$\sim 10$~km$^{2}$ in Yakutsk \cite{Yak-mu-det}, 1 detector in Telescope
%Array \cite{TA-mu-det} and 7 detectors in the Auger Muons and Infill for
%the Ground Array (AMIGA) experiment at Auger \cite{AMIGA}. Even when the
%full AMIGA array of 61 detectors is installed, it would cover only
%25~km$^{2}$.

In this work, we make an attempt to use the data of the Sydney
University Giant Air-shower Recorder (SUGAR) array, which consisted of 54
muon detectors spread over 70~km$^{2}$, for the study of the muon-excess
problem. The present note reports the first results obtained from
comparison of published UHECR spectra measured by two Southern hemisphere
experiments, SUGAR and Auger. We assume that, since both experiments see
the same sky, the true cosmic-ray (CR) spectra should be identical, and
any differences are due to different reconstruction
methods used. The Auger spectrum is normalized to fluorescence detector
energy measurements, a method which is described as
calorimetric with accuracy of $\sim 14\%$ \cite{PAO1}. We therefore use
the Auger spectrum as a proxy to the true CR spectrum seen from the
Southern hemisphere. The SUGAR spectrum was derived from the muon-detector
data, and model assumptions have necessarily been invoked to estimate the
primary energy $E$ from the effective vertical muon number $N_{\rm v}$ for
each shower. In this work, we take the SUGAR $N_{\rm v}$ spectrum and fit
the $E(N_{\rm v})$ relation in such a way that the Auger spectrum is
reproduced from the SUGAR data. Comparison of our empirical $E(N_{\rm v})$
relation with the theoretical one, $E_{\rm S}(N_{\rm v})$, used by the
SUGAR group, reveals the excess of muons in data with respect to the model
relation: $E(N_{\rm v})<E_{\rm S}(N_{\rm v})$, hence the number of
observed muons in a shower with a given primary energy is larger than
expected from the models. This is in line with the results of the Auger
and Yakutsk arrays on the muon content and opens the possibility to study
the muon excess in more detail with SUGAR data on individual events. While
this more detailed study, tracing the origin of the excess, will be
performed elsewhere, we compare here our empirical $E(N_{\rm v})$ relation
with those predicted by Monte-Carlo simulations performed with modern
hadronic-interaction models. This confirms the muon excess in data with
respect to simulations
and
%,for the first time,
 addresses the energy
dependence of this excess:
the
excess grows very moderately with energy over 1.5 decades in the
primary energy between $10^{17}$ and $10^{18.5}$~eV.

In Sec.~\ref{sec:spec}, we discuss briefly the published spectra of SUGAR
and Auger which we use as the input for this study. We compare the spectra
and derive the empirical muon number -- energy relation in
Sec.~\ref{sec:spectra}. This relation is compared to results of
simulations with three hadronic
interaction models in Sec.~\ref{sec:hadron}. Section~\ref{sec:concl}
contains our brief conclusions and discusses prospects for future work.

\section{The spectra}
\label{sec:spec}

\subsection{SUGAR}
\label{sec:spec:SUGAR}
The SUGAR experiment
was in operation between 1968 and 1979~\cite{SUGAR1, Brownlee, SUGAR3}.
The
array was located near the town of Narrabri in New South Wales, Australia
at latitude $30^{\circ}32^{'}$~S, longitude $149^{\circ}36^{'}$~E and
altitude $\sim$250~m above sea level. The array covered an area of about 70
km$^{2}$ and consisted of 54 underground detector stations. There were no
surface-based detectors. Each detector station had two liquid-scintillator
tanks 50 m apart in the North-South direction,  buried at
the depth varying within $1.5\pm0.3$~m  \cite{Brownlee}.
 The effective area of each scintillator tank was
6.0 m$^{2}$. The threshold energy for detected muons was
$(0.75\pm0.15)\sec\theta_{\mu}$ GeV, where $\theta_{\mu}$ is the zenith
angle of the incident muon. To reconstruct the primary energy, readings of
these muon detectors were used as the input. Model-based relations between
the muon number and the energy were invoked.

Early work of the SUGAR group had been criticised for underestimation of
the photomultiplier afterpulsing effect. However, it was subsequently
taken into account properly. In this work, we use the spectrum presented
in Ref.~\cite{SUGARWinn}, where the afterpulsing effect was correctly
taken into
account. Still, the spectrum was considerably higher than those
obtained by other groups. We believe that this is a manifestation of the
``muon excess''.  Note that in an early work, the SUGAR group did study
the surface component. There was a spark chamber detector placed
between two of the closest array SUGAR stations, triggered by a
coincidence of the two stations \cite{SUGAR-spark}. The correction for
afterpulsing was not implemented at that time, but qualitatively, the
results of that study agree well with those of the present work.

\begin{figure}
\centerline{ \includegraphics[width=0.75\linewidth]{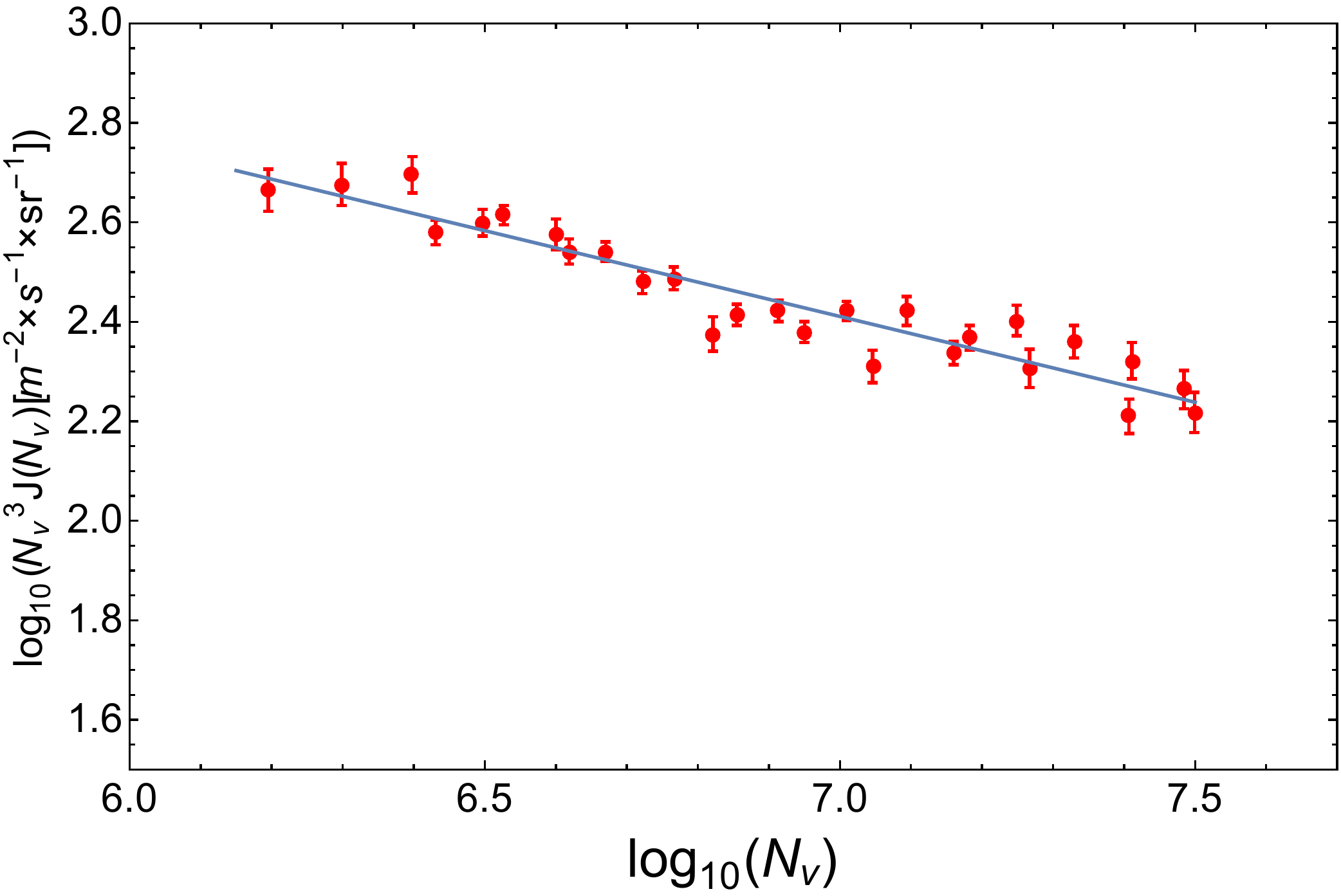}}
\caption{SUGAR differential vertical-equivalent muon-number
($N_{\rm v}$) spectrum~\cite{SUGARWinn}. The line presents the power-law
fit for $N_{\rm v}<10^{7.5}$. }
\label{fig:Flux_Nv}
\end{figure}

We restrict ourselves to the energy range where SUGAR had sufficient
statistics, that is, far below the Greizen--Zatsepin--Kuzmin
suppression. We use the differential vertical muon number, $N_{\rm v}$,
spectrum obtained by combining various muon-number spectra at different
zenith angles \cite{SUGARWinn}. For a given EAS zenith angle $\theta$, the
effective vertical muon number $N_{\rm v}$ in a shower is related to the
reconstructed muon number $N_{\mu }$ through the following relation,
\begin{equation}
\log_{10}\left(\frac{N_{\rm v}}{N_{\rm r}}\right)=
\frac{(1-\gamma_{\rm v}-A(\cos\theta-1))
\log_{10}\left(\frac{N_{\mu}}{N_{\rm r}}\right)+B(\cos{\theta}-1)+\log_{10}
\left(\frac{1-\gamma_{\rm
v}}{1-\gamma_{\rm v}-A(\cos\theta-1)}\right)}{1-\gamma_{\rm v}},
\label{Eq:Nv_Nmu}
\end{equation}
where the coefficients are $A=0.47$, $B=2.33$,
%$E_{r}=1.64\times10^{18}$~eV.
$\gamma_{\rm v}=3.35$, and the normalization scale is $N_{\rm r}=10^{7}$.
This relation was obtained in Ref.~\cite{SUGAR_Nv} empirically from the
data by means of the constant-intensity cuts method. The muon number,
$N_{\mu}$, is, in turn, determined by fitting individual detector readings
by the experimentally determined muon lateral distribution function (LDF),
\begin{equation}
\rho_\mu(r) = N_{\mu}\,k(\theta)\, \left(\frac{r}{r_0}\right)^{-a}\,\left(
1 + \frac{r}{r_0}\right)^{-b}\,\,.
\label{Eq:LDF}
\end{equation}
Here, $\rho_{\mu }$ is the muon density, $N_{\mu}$ is the estimated total
number of muons, $\theta$ is the incident zenith
angle, $r$ is the perpendicular distance from the shower axis, $r_0 =
320$~m, $a = 0.75$, $b= 1.50 + 1.86 \cos\theta$, and
\begin{equation}
k(\theta) = \frac{1}{2\pi r_0^2}\,
\frac{\Gamma(b)}{\Gamma(2-a)\,\Gamma(a+b-2)}\,.
\end{equation}
The resulting $N_{\rm v}$ spectrum, shown in Fig.~\ref{fig:Flux_Nv},
is well described by a power law for $N_{\rm v}<10^{7.5}$, the range used
in this study. More details on the SUGAR data processing may be found in
Ref.~\cite{SUGARWinn}.

According to the SUGAR procedure,
the primary energy of the shower is related to the number of vertical muons
by the following expression,
\begin{equation}
E=E_{\rm r}(N_{\rm v}/N_{\rm r})^{\alpha}.
\label{Eq:E}
\end{equation}
For the Hillas model \cite{H} used in Ref.~\cite{SUGARWinn},
\begin{equation}
E_{\rm r}=1.64\times10^{18}~\mbox{eV},~~~\alpha=1.075.
\label{Eq:1a}
\end{equation}
Given the power-law shape of the $N_{\rm v}$ spectrum and the
relation~(\ref{Eq:E}), the resulting SUGAR energy
spectrum~\cite{SUGARWinn}, see Fig.~\ref{fig:Flux_E} (red empty
triangles),
is also power-law.  The systematic error of the SUGAR
$N_{\rm v}$ spectrum measurement is $\sim 3\%$ as one may conclude from
Ref.~\cite{SUGAR3} (the dominant source of uncertainties discussed there
is related to the determination of the shower geometry). This relatively low
value is not surprising because the dominant source of systematics in the
\textit{energy} spectrum comes from relating the primary energy to
observable quantities, while here we use the muon number measured directly
at the surface and keep the $E(N_{\rm v})$ relation free, see
Sec.~\ref{sec:spectra}. 

\subsection{The Pierre Auger Observatory}
\label{sec:spec:PAO}
The Pierre Auger Observatory has been in operation since 2004 \cite{PAO1}.
The experiment is located near the town of Malarg\"{u}e in Mendoza Province, Argentina at latitude
$35^{\circ}12^{'}$~S, longitude $69^{\circ}19^{'}$~W, and the average
altitude $\sim$1400~m above sea level. The Observatory includes the large
surface-detector (SD) array of 1600 water-Cherenkov detectors
distributed over the area of $\sim 3000$~km$^{2}$, supplemented by an
additional, more dense array of 61 detectors spreading over the area of
$\sim 23.5$~km$^{2}$. Jointly with SD, the fluorescence detector (FD)
consisting of 4 telescope stations works in coincidence with the large SD
during a limited fraction of time (clear, moonless nights). An additional
FD station, the High-Elevation Auger Telescope (HEAT), is able to detect
lower-energy showers in coincidence with the dense SD array. The FD method
 provides the primary energy estimates with reduced model
uncertainties and is therefore used to calibrate the SD energy scale by
means of simultaneous (hybrid) observations of a number of events.
The systematic error of the Auger energy scale is $\pm
14\%$ \cite{PAO1} and is dominated by the uncertainty in the
fluorescence yield measurement. For the present study, we use the most
recent combined energy spectrum reported in Ref.~\cite{Aab:2017njo}. In
the energy range we use, $E \lesssim 10^{18.5}$~eV, the spectrum is
dominated by the dense SD data, with a modest contribution from hybrid
data in the higher-energy part, and is well approximated by a power
law~\cite{Aab:2017njo}. The spectrum is also shown in
Fig.~\ref{fig:Flux_E}  (blue filled circles).

\section{Comparison of the spectra}
\label{sec:spectra}
\begin{figure}
\centerline{
\includegraphics[width=0.75\linewidth]{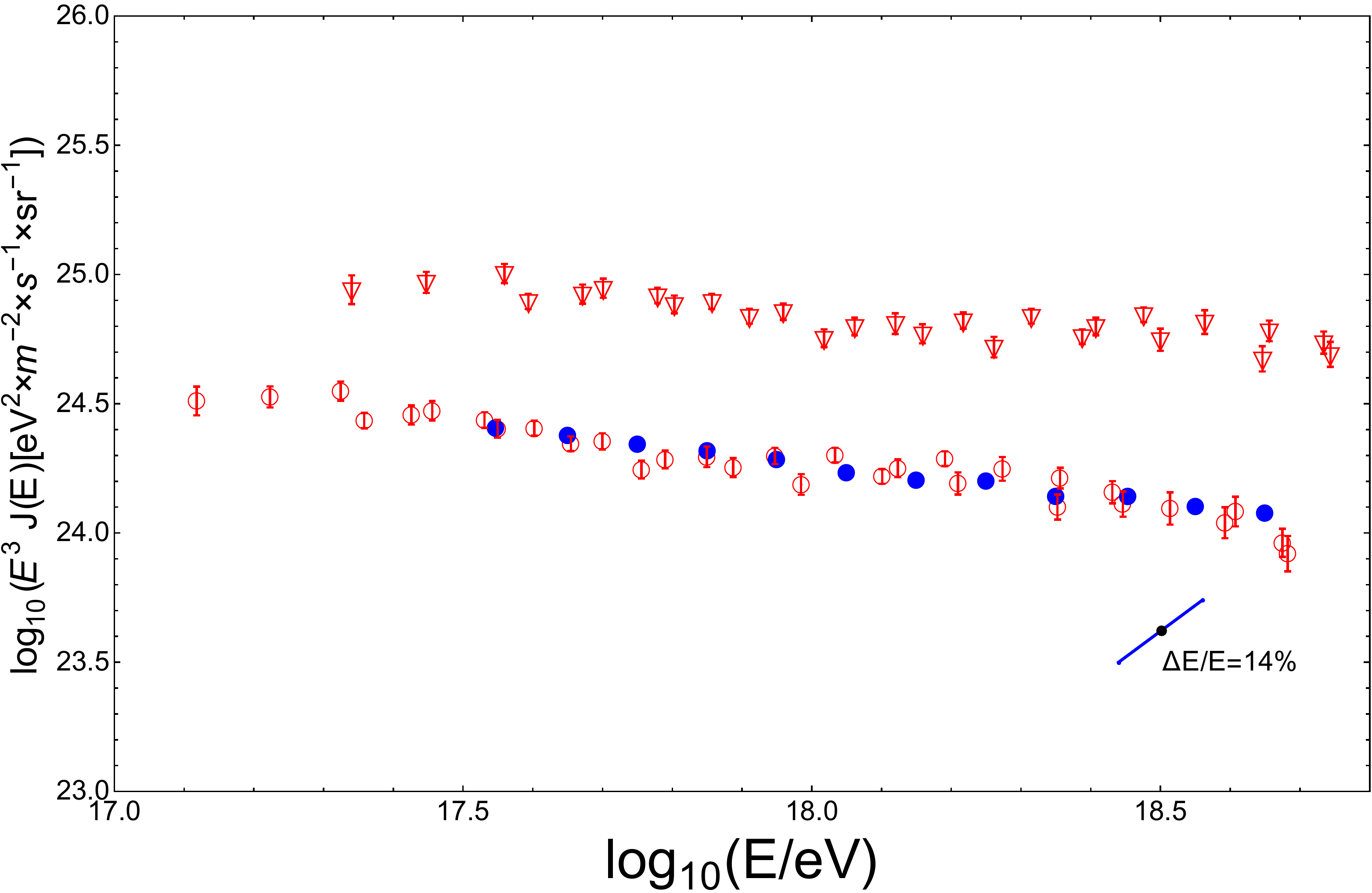}}
\caption{The SUGAR differential energy spectrum~\cite{SUGARWinn} estimated
using the Hillas model \cite{H} in Eqs.~(\ref{Eq:E}) and (\ref{Eq:1a})
(red
empty triangles); the Auger differential combined energy spectrum
\cite{Aab:2017njo} (blue filled circles); the SUGAR differential energy
spectrum  estimated using the empirical $E(N_{\rm v})$ relation in
Eqs.~(\ref{Eq:E}) and (\ref{Eq:6*}) (red empty circles; this work).
The inclined bar represents the systematic uncertainty in terms of $E$,
summed in quadratures.}

\label{fig:Flux_E}
\end{figure}

The key part of the present study is to change the $E(N_{\rm v})$ relation
in such a way that the SUGAR muon data produces the Auger spectrum
calibrated by the FD. This is straightforward to do given the power-law
shapes of both spectra. We fit parameters $E_{\rm r}$ and $\alpha$ of
Eq.~(\ref{Eq:E}) requiring that the SUGAR energy spectrum, reconstructed
with these new parameters, matches the Auger combined energy spectrum. As
a result we obtain the following values of parameters for
Eq.~(\ref{Eq:E}),
%\begin{equation}
\begin{eqnarray}
E_{\rm r}=(8.67 \pm 0.21_{\rm stat} \pm 0.26_{\rm
syst~SUGAR}\pm1.21_{\rm syst~Auger})\times10^{17}~\mbox{eV},
\nonumber \\
\alpha=1.018\pm 0.0042_{\rm stat}\pm 0.0043_{\rm
syst~SUGAR}\pm0.0028_{\rm syst~Auger},
\label{Eq:6*}
\end{eqnarray}
%\end{equation}
 where the systematics discussed above was propagated. 
The modified SUGAR spectrum is shown in Fig.~\ref{fig:Flux_E} by red empty
circles and demonstrates an excellent agreement with the Auger spectrum.
\section{Comparison to the Monte-Carlo simulations}
\label{sec:hadron}
At the next step, we quantify the ``muon excess'' by comparing our
empirical $E(N_{\rm v})$ relation with those predicted by modern
Monte-Carlo (MC) models. We  use  the CORSIKA 7.4001 \cite{Heck:1998vt}
EAS
simulation package. We choose the QGSJET-II-04 \cite{Ostapchenko:2010vb},
EPOS-LHC \cite{EPOS-LHC} and SYBYLL-2.3c \citep{SIBYLL} as the high-energy
hadronic interaction
models and FLUKA2011.2c \cite{Fluka} as the low-energy hadronic
interaction model. We simulate a library of artificial EAS with
primary energies following an $E^{-3.19}$ differential spectrum
\cite{SUGARWinn} with
$9\times10^{16}$~eV$<E<4\times10^{18}$~eV. These EAS are
simulated with zenith angles in the range between $0^{\circ}$ and
$75^{\circ}$ (as in the real SUGAR data used in Ref.~\cite{SUGARWinn})
assuming an isotropic distribution of arrival directions in the celestial
sphere. The simulations were performed with the thinning parameter
$\epsilon= 10^{-5}$ and maximal weight limitations,
cf.\ Ref.~\cite{Kobal}.  As it is customary in simulations with
thinning for sparse ground arrays, particles within 100~m from the
core were discarded.

For each of the three high-energy hadronic
interaction models, we simulated 10000 showers for primary protons and the
same number of showers for primary iron. The lower energy thresholds are
fixed for hadrons (excluding $\pi^{0}$) and muons as 50~MeV; for photons,
e$^{+}$, e$^{-}$ and $\pi^{0}$ as 250~keV. The standard geomagnetic field
for the SUGAR array location is assumed, $B_{x}=24.0~\upmu$T and
$B_{z}=-51.4~\upmu$T.

Since we are interested in the mean $E(N_{\rm v})$ relation and not in its
fluctuations, we estimate the number of muons with energies above the
SUGAR threshold in
an artificial shower
and calculate the muon density in concentric rings around the shower axis.
As was done with the real data, we use the experimental
muon LDF in Eq.~(\ref{Eq:LDF}) and fit it to the muon
density distribution in MC, obtaining $N_{\mu}$.
Then, we use Eq.~(\ref{Eq:Nv_Nmu}) to
express the effective number of vertical muons $N_{\rm v}$ through
$N_{\mu}$ and $\theta$. As a result, $N_{\rm v}$ is determined for each
artificial shower. We note that this procedure gives, of course, a rough
approximation to the data analysis chain, since, in reality, only a few
detector stations were firing, which reduced considerably the precision of
reconstruction of the core position, the arrival angles and the LDF.
 To demonstrate that mean relations are not strongly affected by these
uncertainties, we performed a simple simulation in which the effect of a
limited number of stations was imitated in the following way. A random
location of the shower within the SUGAR array was generated and the mean
muon density at each detector station was calculated from LDF and
transformed into detector readings by means of the Poisson distribution.
The true geometry was modified to imitate reconstruction errors (50~m in
the core location and 2.6$^{\circ}$ in the arrival direction, assuming
Gaussian distributions). Then, $N_{\mu }$ was reconstructed from these
data and compared to the input one. For an \textit{individual} shower, the
statistical error introduced by this procedure was $\sim 19\%$ in $N_{\mu
}$, \textit{symmetric} with respect to the central value. As a result, the
total effect of our simplified procedure on the reconstruction of the MC
relations from 10000 artiicial events was far below statistical
fluctuations. A more detailed study, taking into account all steps of the
SUGAR reconstruction procedure, will be performed elsewhere. 

\begin{figure}
\centerline{
\includegraphics[width=0.75\linewidth]{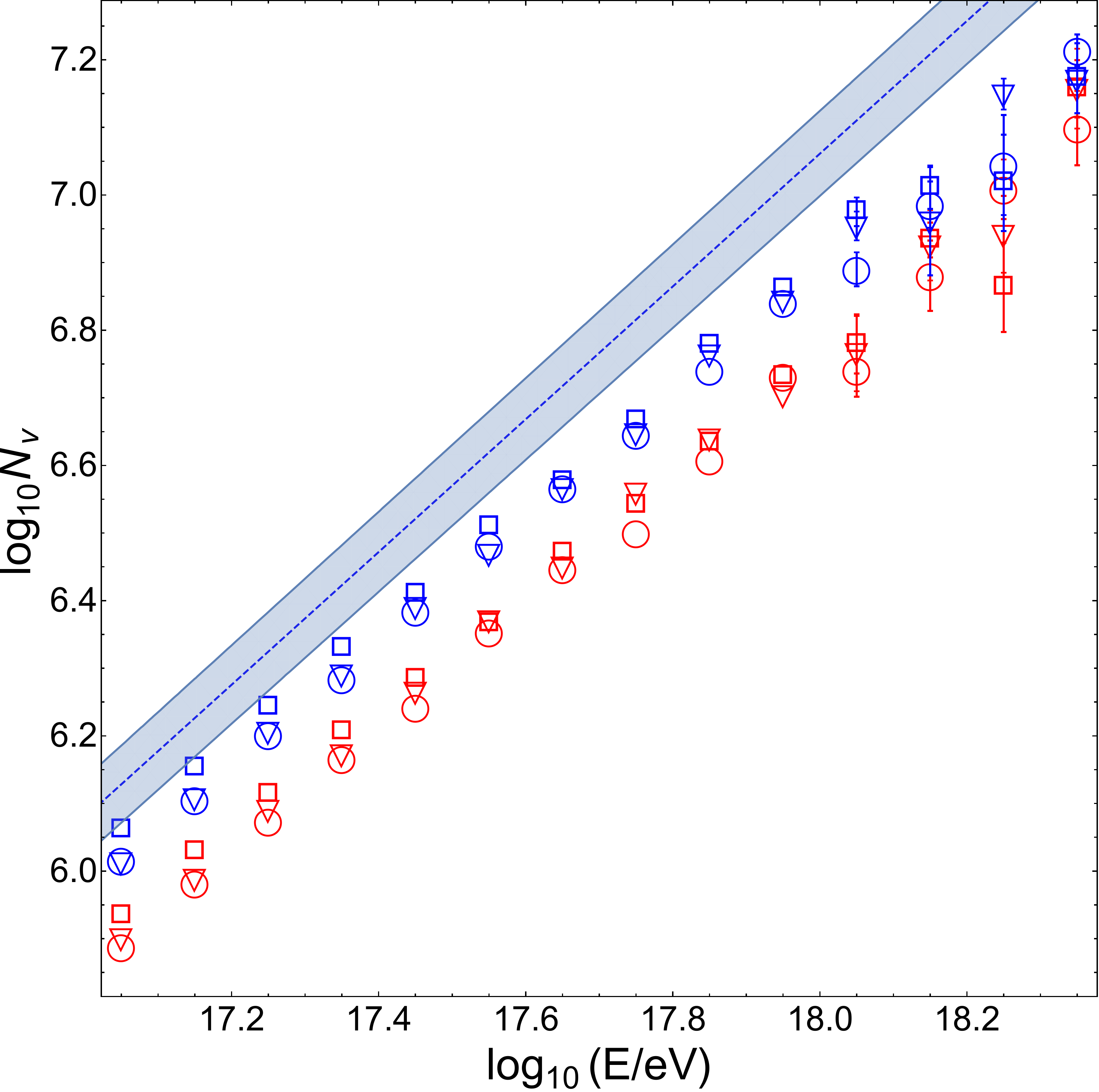}}
\caption{Mean effective number of vertical muons $N_{\rm v}$ as a function
of the primary energy. Points indicate the results of Monte-Carlo
simulations with QGSJET-II-04 (protons - red open circles, iron - blue
open circles), EPOS-LHC (protons - red open triangles, iron - blue
open triangles) and SIBYLL-2.3c (protons - red open squares, iron - blue
open squares).  The dashed blue line
corresponds to our empirical model Eqs.~(\ref{Eq:E}) and (\ref{Eq:6*}); the
shaded blue area indicates  the total uncertainly (statistical and
systematic errors summed in quadrature). Statistical error bars for the
Monte-Carlo points are shown where they are larger than symbols. }
\label{fig:Nv(E)}
\end{figure}

\begin{figure}
\centerline{
\includegraphics[width=0.85\linewidth]{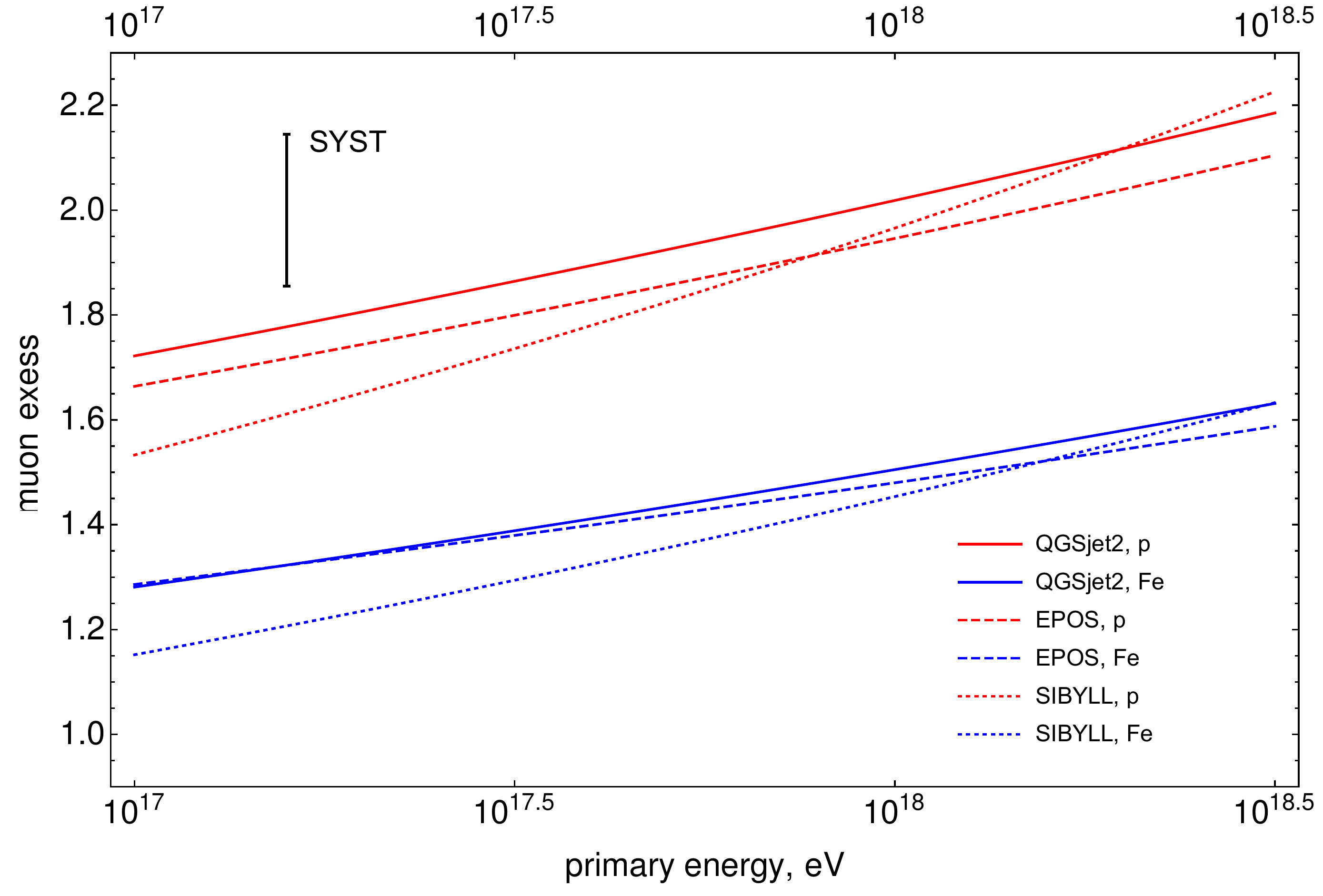}}
\caption{Best-fit muon excess, $N_{\rm v}/N_{\rm v}^{\rm MC}$, as
a function of the primary energy $E$.  QGSJET-II-04 -- full lines,
% for
%central values and shaded regions for error bars (protons - red, iron -
%blue);
EPOS-LHC -- dashed lines;
SIBYLL-2.3c -- dotted lines
(protons - red, iron - blue). The vertical bar indicates the estimated
systematic error.  }
\label{fig:fig4}
\end{figure}

Figure~\ref{fig:Nv(E)}
presents a comparison of the simulation results for $N_{\rm v}(E)$ with our
empirical relation. The ``muon excess'' is clearly seen, and its
dependence on the energy may be studied. To this end, we parametrise the
ratio of empirical, $N_{\rm v}^{\rm emp}$, and simulated, $N_{\rm v}^{\rm
MC}$, muon numbers  in vertical showers  as a power law in energy
$E$,
\begin{equation}
 \frac{N_{\rm v}}{N_{\rm v}^{\rm MC}}    =
\left(
\frac{N_{\rm v}}{N_{\rm v}^{\rm MC}}
   \right)_{0}
\left(\frac{E}{E_{0}}   \right)^{q} .
\label{Eq:A*}
\end{equation}
\black
Best-fit normalizations and exponents for the relation (\ref{Eq:A*}) are
given in Table~\ref{Tab:1}
\begin{table}
\begin{tabular}{ccc}
\hline
Simulation & $\left(N_{\rm v}/N_{\rm v}^{\rm MC}\right)_{0}$ &
$q$ \\
\hline
\hline
QGSJET-II-04 protons ~ & $1.722\pm  0.036_{\rm stat}\pm 0.253_{\rm
syst}$ \black & ~~ $0.069\pm  0.016_{\rm stat}\pm 0.007_{\rm syst}$ \\
QGSJET-II-04 iron ~ & $1.281\pm  0.011_{\rm stat}\pm 0.188_{\rm syst}$  & ~~
$0.070\pm  0.006_{\rm stat}\pm 0.007_{\rm syst}$ \\
EPOS-LHC protons ~ & $1.664\pm  0.027_{\rm stat}\pm 0.244_{\rm syst}$  & ~~
$0.068\pm  0.012_{\rm stat}\pm 0.007_{\rm syst}$ \\
EPOS-LHC iron ~ & $1.285\pm  0.013_{\rm stat}\pm 0.189_{\rm syst}$  & ~~
$0.061\pm  0.008_{\rm stat}\pm 0.007_{\rm syst}$ \\
SIBYLL-2.3c protons ~ & $1.533\pm 0.014_{\rm stat}\pm 0.225_{\rm syst}$  &
~~ $0.108\pm  0.007_{\rm stat}\pm 0.007_{\rm syst}$ \\
SIBYLL-2.3c iron ~ & $1.152\pm  0.015_{\rm stat}\pm 0.169_{\rm syst}$  & ~~
$0.101\pm  0.010_{\rm stat}\pm 0.007_{\rm syst}$ \\
\hline
\end{tabular}
\caption{\label{Tab:1}
Normalizations and exponents of the energy-dependent muon excess relation,
Eq.~(\ref{Eq:A*}), for $E_{0}=10^{17}$~eV.}
\end{table}
for $E_{0}=10^{17}$~eV and various models and primaries. These relations,
quantifying the muon excess, are plotted in Fig.~\ref{fig:fig4}.
We see that the muon excess
grows moderately with energy. It could be possible that extrapolation of
this dependence to even lower energies may result in the disappearence of
the excess at $E_{0} \sim 10^{16}$~eV.
%, as suggested by some observations
%in this energy range~\cite{IceTop}.
We note that the discrepancy between
the data and simulations is less pronounced for heavier primaries, though
still significant. Agreement at $E_{0}\sim 10^{17}$~eV between the three
hadronic-interaction models we use is not surprising because  all
three  were tuned to reproduce the Large Hadron Collider results.

\section{Conclusions and outlook}
\label{sec:concl}
In this paper, we obtained an empirical relation between the number of
muons in an extensive air shower and the primary energy, for energies
$10^{17}$~eV$\lesssim E \lesssim 10^{18.5}$~eV. To this end, we took the
SUGAR effective muon-number spectrum and determined the
 relationship (\ref{Eq:E}), (\ref{Eq:A*})
between the energy and muon number in a vertical shower, starting 
from the requirement that the
Auger energy spectrum is reproduced from the SUGAR muon data. Then, we
compared our empirical relation with the Monte-Carlo simulations performed
with the help of modern hadronic-interaction models. We found the excess
of muons in real air showers with respect to simulations, as parametrized
by Eq.~(\ref{Eq:A*}) with parameters listed in Table~\ref{Tab:1}, for three
hadronic-interaction models, QGSJET-II-04, EPOS-LHC and SIBYLL-2.3c, and two types of
primary particles, protons and iron nuclei.
%The energy dependence of the muon excess has never been studied before in the frameworks of a single data set

In our work, we treated the Auger energy spectrum as the true underlying
primary spectrum for the SUGAR events. This assumption is motivated by the
similar fields of view of the two Southern experiments and by the
calibration of the Auger spectrum to the FD measurements, which suppresses model dependencies. Nevertheless, the correctness of
this assumption remains the main source of the systematic uncertainties of
our approach, and the systematic error of the
experimental energy scale dominates the uncertainties of
our study.

This work represents the first step in exploitation of unique SUGAR muon
data for a detailed study of the muon excess. The shower-by-shower SUGAR
data will be analized to compare real and simulated muon numbers as a
function of not only the primary energy, but also of the atmospheric depth
(zenith-angle dependence), distance to the shower core (the LDF shape)
etc., allowing us to trace the origin of the discrepancies between data
and simulations. This work, currently in progress, would ultimately affect
both our understanding of hadronic interactions and the interpretation of
all high-energy cosmic-ray results.

\section*{Acknowledgements}
We thank
Leonid Bezrukov,
Bruce Dawson,
Oleg Kalashev, Mikhail Kuznetsov, Maxim Pshirkov, Peter
Tinyakov and Yana Zhezher for helpful discussions. Monte-Carlo simulations
have been performed at the computer cluster of the Theoretical Physics
Department, Institute for Nuclear Research of the Russian Academy of
Sciences. Development of the analysis methods (IK and GR) is supported by
the Russian Science Foundation (grant 17-72-20291).

\end{document}